\documentclass[preprintnumbers,nofootinbib,amsmath,amssymb]{revtex4}

\usepackage{color} 
     
\usepackage{amssymb}
\usepackage{amsmath}
\usepackage{graphicx}

\begin{document}

\title{Magnetic Charge and Photon Mass: Physical String Singularities, Dirac Condition, and Magnetic Confinement}

\author{Timothy J. Evans}
\email{tevans559@mail.fresnostate.edu}
\affiliation{Department of Physics, California State University Fresno, Fresno, CA 93740-8031, USA}

\author{Douglas Singleton}
\email{dougs@mail.fresnostate.edu}
\affiliation{Department of Physics, California State University Fresno, Fresno, CA 93740-8031, USA \\
and \\
Institute of Experimental and Theoretical Physics Al-Farabi KazNU, Almaty, 050040, Kazakhstan}

\date{\today}

\begin{abstract}
We find exact, simple solutions to the Proca version of Maxwell's equations with magnetic sources. Several properties of these solutions differ from the ususal case of magnetic charge with a massless photon: (i) the string singularities of the usual 3-vector potentials become real singularities in the magnetic fields; (ii) the different 3-vector potentials become gauge inequivalent and physically distinct solutions; (iii) the magnetic field depends on $r$ and $\theta$ and thus is no longer rotationally symmetric; (iv) a combined system of electric and magnetic charge carries a field angular momentum even when the electric and magnetic charges are located at the same place ({\it i.e.} for dyons); (v) for these dyons one recovers the standard Dirac condition despite the photon being massive. We discuss the reason for this. We conclude by proposing that the string singularity in the magnetic field of an {\it isolated} magnetic charge suggests a confinement mechanism for magnetic charge, similar to the flux tube confinement of quarks in QCD.             
\end{abstract}

%\pacs{14.80.Hv, 11.27.+d, 14.70.Bh. \\ Keywords: Magnetic Charge, Photon Mass, Magnetic Confinement}

\maketitle
 
\section{Introduction}

The incorporation of magnetic charge into Maxwell's equations taking into account basic quantum mechanics begins with the works of Dirac \cite{dirac} \cite{dirac1} which showed that allowing magnetic charge gave a quantization condition between electric and magnetic charges. The basic set up Dirac considered was a Coulomb magnetic field ${\bf B} = g \frac{{\bf {\hat r}}}{r^2}$ with magnetic charge $g$ coming from a 3-vector potentials of the form   
\begin{equation}
\label{A-string}
{\bf A} ^{(0)} _{\pm} = \frac{g}{r \sin \theta} (\pm 1 - \cos \theta ) {\bf {\hat \varphi}}~.
\end{equation}   
The superscript stands for zero photon mass. The two potentials in \eqref{A-string} are singular along the entire $+z$-axis (${\bf A} ^{(0)} _-$) or $-z$-axis (${\bf A} ^{(0)} _+$). These string singularities are are not inherited by the magnetic field which has only the usual point source singularity at $r=0$. Also ${\bf A} ^{(0)} _+$ and ${\bf A} ^{(0)} _-$ are related via a gauge transformation so these different potentials lead to the same magnetic field. Further one can hide the string singularity in the gauge potential, making it unphysical, by requiring that the Aharonov-Bohm phase shift \cite{jackson} \cite{ab} is some integer multiple of $2 \pi$. This leads to the famous Dirac quantization condition between electric ($e$) and magnetic ($g$) charges
\begin{equation}
\label{dirac-cond}
e g = n \frac{\hbar}{2} ~.
\end{equation}
We set $c=1$ throughout. This condition can also be obtained by calculating the field angular momentum of an electric charge and magnetic charge separated a vector ${\bf R}$ \cite{saha} \cite{saha1} \cite{wilson} giving  
\begin{equation}
\label{ang3d}
{\bf L}_{EM} ^{(0)} = \frac{1}{4 \pi} \int {\bf r} \times ({\bf E} \times {\bf B}) d^3 x = - e g {\bf {\hat R}}~,
\end{equation} 
where ${\bf {\hat R}}$ is the unit vector from the magnetic charge to the electric charge. Fixing the magnitude of ${\bf L}_{EM} ^{(0)}$ to be some integer multiple of $\frac{\hbar}{2}$ immediately gives \eqref{dirac-cond}. We now investigate how allowing the photon to become massive changes some of these results. \\

\section{Maxwell's equations with magnetic charge and non-zero photon mass}

For a photon with a mass $m$, Maxwell's equations with both electric and magnetic sources are (in Gaussian units and using 4-vector notation)
\begin{equation}
\label{maxwell4-source}
\partial _\nu F^{\nu \mu} + m ^2 A^\mu = 4 \pi J^\mu _{(e)} ~~~~;~~~~ \partial _\nu {\cal F}^{\nu \mu} = 4 \pi J^\mu _{(m)} ~,
\end{equation}     
where $F^{\mu \nu} = \partial ^\mu A^\nu - \partial^\nu A^\mu$ and ${\cal F}^{\mu \nu} = \frac{1}{2} \epsilon^{\mu \nu \alpha \beta} F_{\alpha \beta}$ are the usual field strength tensor and dual field strength tensor respectively. With the addition of the mass term, $m ^2 A^\mu$, the equations in \eqref{maxwell4-source} are known as the Proca equations. The equations given in \eqref{maxwell4-source} are in the 4-vector form written down in reference \cite{joshi} in order to study magnetic charge with a photon mass term (compare \eqref{maxwell4-source} to equations (3a)(3b) in \cite{joshi}) . Looking at the two equations in \eqref{maxwell4-source} one can see that adding the photon mass term, $m ^2 A^\mu $, spoils the usual duality between electric and magnetic quantities of Maxwell's equations with a massless photon. 

There are other approaches to magnetic charge where one can extend the duality down to the level of the potentials. The best known, alternative approach is the Cabibbo-Ferrari formulation of magnetic charge \cite{cabibbo} where one introduces two 4-vector potentials. A detailed discussion of how duality and magnetic charge co-exist in the Cabibbo-Ferrari formulation can be found in references \cite{zwanzinger} \cite{singleton} \cite{singleton1}. In addition, reference \cite{singleton} \cite{singleton1} incorporates a photon mass term in the Cabibbo-Ferrari formulation of magnetic charge using the Higgs mechanism rather than adding the photon mass term by hand as was done in \eqref{maxwell4-source} and in \cite{joshi}. In this way one breaks the initial $U(1) \times U(1)$ symmetry down to a single $U(1)$ symmetry, with one massless photon (the ``electric" photon) and one massive photon (the ``magnetic" photon). There are other works which also consider an expanded potential sector. Reference \cite{ahrens} is the earliest work we found to consider both magnetic charge and photon mass. This paper also introduced two 4-vector potentials but which differ in detail from the Cabibbo-Ferrari formulation. Reference \cite{neto} also considered two gauge potentials, but followed the Cremmer-Scherk-Kalb-Ramond (CSKR) model \cite{cremmer} \cite{kalb} where one gauge potential was a 4-vector while the other gauge potential was a 4-tensor of rank two. Both \cite{ahrens} and \cite{neto} look at the effect of combining magnetic charge plus a mass term for one or more of the gauge potentials.    

In this paper we follow the step up of reference \cite{joshi} where there is only one 4-vector potential and with the photon mass term put in by hand. This means we are giving up electric-magnetic duality. The 4-vector form of the equations given in \eqref{maxwell4-source} can be written in 3-vector form as 
\begin{eqnarray}
\label{e-maxwell3-source}
&&\nabla \cdot {\bf E} + m ^2 \phi = 4 \pi \rho _{(e)}  \\
\label{e-maxwell3-source-1}
&&\nabla \times {\bf E} + \frac{\partial {\bf B}}{\partial t} = - 4 \pi {\bf J}_{(m)}  \rightarrow \nabla \times {\bf E} = 0     \\ 
\label{b-maxwell3-source}
&& \nabla \cdot {\bf B} = 4 \pi \rho _{(m)}  \\
\label{b-maxwell3-source-1}
&& \nabla \times {\bf B} - \frac{\partial {\bf E}}{\partial t} + m ^2 {\bf A} = 4 \pi {\bf J}_{(e)}  \rightarrow 
\nabla \times {\bf B} + m ^2 {\bf A} = 0 ~.
\end{eqnarray}
The system of equations \eqref{e-maxwell3-source} \eqref{e-maxwell3-source-1} \eqref{b-maxwell3-source} \eqref{b-maxwell3-source-1} are those given in \cite{joshi}. We will find exact, simple solutions to equations \eqref{e-maxwell3-source} \eqref{e-maxwell3-source-1} \eqref{b-maxwell3-source} \eqref{b-maxwell3-source-1} which differ from those found in \cite{joshi}. The solution we find here is different from those found in the two potential approaches of \cite{singleton} \cite{singleton1} \cite{ahrens} \cite{neto}. 

We now present an exact, closed form solution to equations \eqref{e-maxwell3-source} \eqref{e-maxwell3-source-1} \eqref{b-maxwell3-source} \eqref{b-maxwell3-source-1}. At the level of the potentials one has ${\bf E} = -\nabla \phi - \frac{\partial {\bf A}}{\partial t} \rightarrow -\nabla \phi$ and ${\bf B} = \nabla \times {\bf A}$. The arrows in \eqref{e-maxwell3-source-1} \eqref{b-maxwell3-source-1} indicate the limit we are considering of time-independent fields ($\partial _t {\bf E} = \partial _t {\bf B} = 0$) and static sources (${\bf J}_{(e)} = {\bf J}_{(m)} = 0 $). The expression ${\bf E} = -\nabla \phi$ can be combined with the divergence of the electric field in \eqref{e-maxwell3-source} to give $-\nabla ^2 \phi + m ^2 \phi = 4 \pi \rho _{(e)}$. For a delta function source $\rho _{(e)} = e \delta ^3 ({\bf r})$ this has the well known Yukawa solution
\begin{eqnarray}
\label{e-yukawa}
&&\phi = e \frac{e^{-m r}}{r} \\ 
\label{e-yukawa-1}
&&{\bf E} = -\nabla \phi =e \left( 1 + m r \right) \frac{e^{-m r}}{r^2} {\bf {\hat r}} ~.
\end{eqnarray}
In the previous study of solutions to the system of equations \eqref{e-maxwell3-source} \eqref{e-maxwell3-source-1} \eqref{b-maxwell3-source} \eqref{b-maxwell3-source-1} reference \cite{joshi} used ${\bf E} = e \frac{e^{-m r}}{r^2} {\bf {\hat r}}$ for their electric field, whereas our electric field is given by \eqref{e-yukawa-1}. We believe that \cite{joshi} incorrectly applied the Yukawa form to the field rather than the scalar potential as given in \eqref{e-yukawa} above. It is straightforward to check that $\phi , {\bf E}$ from \eqref{e-yukawa} \eqref{e-yukawa-1} solve the electric field equations \eqref{e-maxwell3-source} \eqref{e-maxwell3-source-1} and as well $\phi$ solves $-\nabla ^2 \phi + m ^2 \phi = 4 \pi e \delta ^3 ({\bf r})$. Thus equations \eqref{e-yukawa} and \eqref{e-yukawa-1} are exact solutions to the electric half of the Proca equations with electric point sources. Equations \eqref{e-yukawa} and \eqref{e-yukawa-1} are the well known Yukawa solution for the scalar potential and associated electric field. We now move on to the two magnetic field equations \eqref{b-maxwell3-source} \eqref{b-maxwell3-source-1} and show that there is a simple, Yukawa-like solution to these equations. 

Inserting ${\bf B} = \nabla \times {\bf A}$ into the second equation in \eqref{b-maxwell3-source} yields the second order equation $-\nabla ^2 {\bf A} + m ^2 {\bf A} = 0$ which is similar to the second order equation for $\phi$ but without the source term. \footnote{To arrive at this result one uses $\nabla \cdot {\bf A}=0$ which is {\it not} the Coulomb gauge condition. In massive electrodynamics one requires $\partial _\mu A^\mu =0$ by current conservation \cite{jackson} which since we are considering time-independent potentials and fields becomes $\nabla \cdot {\bf A}=0$.} In the electric case the scalar potential for a massive photon was obtained simply by multiplying the Coulomb potential by $e^{-mr}$ giving $\phi$ in \eqref{e-yukawa}. Motivated by the electric case, we assume that the massive 3-vector potential is given by ${\bf A}^{(0)} _\pm$ multiplied by $e^{-mr}$. From this assumed form for ${\bf A}_\pm$ we obtain the magnetic field via ${\bf B} _\pm = \nabla \times {\bf A}_\pm$.
\begin{eqnarray}
\label{b-yukawa}
&&{\bf A}_\pm = e^{-mr} {\bf A}^{(0)} _\pm (r) = g \frac{e^{-m r}}{r} \left( \frac{\pm 1 - \cos \theta }{\sin \theta} \right) {\bf {\hat \varphi}} \\
\label{b-yukawa-1}
&&{\bf B} _\pm = g \frac{e^{-m r}}{r^2} \left( {\bf {\hat r}} + m r \left[\frac{\pm 1 - \cos \theta }{\sin \theta} \right] {\bf {\hat \theta}} \right) ~.
\end{eqnarray}
Here ${\bf A} ^{(0)} _\pm (r)$ is 3-vector potential in the massless case \eqref{A-string}.
We will now show that \eqref{b-yukawa} \eqref{b-yukawa-1} \footnote{The magnetic field in \eqref{b-yukawa-1} is obtained by taking the curl of  
${\bf A}_\pm = e^{-mr} {\bf A}^{(0)} _\pm (r)$. ${\bf B}_\pm = \nabla \times (e^{-mr} {\bf A}^{(0)} (r)) = e^{-mr} (\nabla \times {\bf A}^{(0)} _\pm (r)) + (\nabla e^{-mr} ) \times {\bf A} ^{(0)}_\pm (r)$. Since $\nabla e^{-mr} = - m e^{-m r} {\bf {\hat r}}$ and $\nabla \times {\bf A}^{(0)} _\pm (r) = g \frac{{\bf {\hat r}}}{r^2}$ we arrive at \eqref{b-yukawa-1}.} are exact solutions to the two magnetic equations \eqref{b-maxwell3-source} \eqref{b-maxwell3-source-1}, as well as satisfying $-\nabla ^2 {\bf A} + m ^2 {\bf A} = 0$ and the requirement $\nabla \cdot {\bf A} =0$. 

First, since ${\bf A}_\pm$ has only a ${\hat \varphi}$ component and does not depend on $\varphi$ one immediately sees that $\nabla \cdot {\bf A}_\pm =0$. Second, using $\nabla \times {\bf B}_\pm = - m^2 g \frac{ e^{-m r}}{r} \left( \frac{\pm 1 - \cos \theta }{\sin \theta} \right) {\bf {\hat \varphi}}$ one can see that $\nabla \times {\bf B}_\pm + m ^2 {\bf A}_\pm = 0$ is satisfied. Third, for the Helmholtz-like equation $-\nabla ^2 {\bf A} + m ^2 {\bf A} = 0$ we need to take into account that $\nabla ^2 {\bf A} \rightarrow (\nabla^2 A_\varphi - \frac{A_\varphi}{r^2 \sin ^2 \theta}) {\bf {\hat \varphi}}$ where $\nabla ^2 = \frac{1}{r^2} \frac{\partial}{\partial r}\left( r^2 \frac{\partial}{\partial r}\right) +\frac{1}{r^2 \sin \theta} \frac{\partial}{\partial \theta} \left( \sin \theta \frac{\partial}{\partial \theta}\right) +\frac{1}{r^2 \sin ^2 \theta} \frac{\partial ^2}{\partial \varphi ^2}$ is the usual Laplacian applied to a scalar quantity. Using ${\bf A}_\pm$ from \eqref{b-yukawa} and performing the lengthy derivatives shows that $-\nabla ^2 {\bf A}_\pm + m ^2 {\bf A}_\pm = 0$. Fourth, for the divergence of the magnetic field we calculate the radial component first as
\begin{eqnarray}
\label{div-b-1}
\nabla \cdot \left( g \frac{e^{-m r}}{ r^2} {\bf {\hat r}} \right) = g e^{-m r} \nabla \cdot \left(\frac{{\bf {\hat r}} }{r^2} \right)
+g \left(\frac{{\bf {\hat r}} }{ r^2} \right) \cdot \nabla (e^{-m r}) = 4 \pi g \delta ^3 ({\bf r}) - g \frac{m e^{-m r}}{r^2}~.
\end{eqnarray}   
We have used $\nabla \cdot \left(\frac{{\bf {\hat r}} }{r^2} \right) = 4 \pi \delta ^3 ({\bf r})$ and 
$e^{-m r} \delta ^3 ({\bf r}) \rightarrow \delta ^3 ({\bf r})$. Evaluating the divergence of the ${\bf {\hat \theta}}$ component of 
${\bf B}_\pm$ gives 
\begin{equation}
\label{div-b-2}
\nabla \cdot \left( g \frac{m e^{-m r}}{r} \left[\frac{\pm 1 - \cos \theta }{\sin \theta} \right] {\bf {\hat \theta}} \right) = 
g \frac{m e^{-m r}}{r^2}~,
\end{equation}
combining \eqref{div-b-1} and \eqref{div-b-2} solves equation \eqref{b-maxwell3-source} with a delta function source {\it i.e.} $\nabla \cdot {\bf B} _\pm = 4 \pi g \delta ^3 ({\bf r})  = 4 \pi \rho _{(m)}$. 

We have shown that the potentials and fields in \eqref{e-yukawa} \eqref{e-yukawa-1} \eqref{b-yukawa} \eqref{b-yukawa-1} are exact solutions for the electric and magnetic fields of Maxwell's equations with both electric and magnetic sources and a photon mass. The electric case in \eqref{e-yukawa} \eqref{e-yukawa-1} is the well known Yukawa solution but the results for the 3-vector potential and magnetic field in \eqref{b-yukawa} \eqref{b-yukawa-1} are given here for the first time as far as we know. All the potentials and fields fall off with the factor $e^{-m r}$ as one would expect for a massive photon. This is in contrast to the claim for the magnetic field in \cite{joshi} which is built around the usual Dirac string 3-vector potential in \eqref{A-string} and thus has a Coulomb magnetic field in the far field region. Physically this is not what one would expect -- with a photon mass term the fields and potentials are expected to fall off like $e^{-m r}$ as we found for the solutions presented here. In any case by explicit calculation we have shown that \eqref{e-yukawa} \eqref{e-yukawa-1} \eqref{b-yukawa} \eqref{b-yukawa-1} are exact solutions to the Proca equations with point electric and magnetic sources {\it i.e.} \eqref{e-maxwell3-source} \eqref{e-maxwell3-source-1} \eqref{b-maxwell3-source} \eqref{b-maxwell3-source-1}. \\

\noindent There are several noteworthy aspects to the magnetic field solution in \eqref{b-yukawa-1}: \\

(i) The string singularity of the 3-vector potential is still present in ${\bf B}$ through the ${\bf {\hat \theta}}$ component. By adding a photon mass the string singularity has become a physical object. One can see the physical character of the string singularity in ${\bf B}$ by calculating the magnetic field energy ($ \frac{1}{8 \pi} \int {\bf B}^2 d^3 x$) and see that it diverges not only at $r=0$, as in the massless photon case, but all along the $\pm z$ axis. \\
 
(ii) ${\bf B}_+$ and ${\bf B}_-$ are physically distinct since the string singularities are on different axes. This is different from the massless case where ${\bf A}^{(0)} _\pm$ gave the same Coulomb magnetic field. More over ${\bf A}^{(0)}_+$ and ${\bf A}^{(0)}_-$ are related by the gauge transformation $A ^{(0)} _- =  A ^{(0)} _+ - \nabla (\lambda )$ with $\lambda = 2g \varphi$, thus guaranteeing that the magnetic field coming from either ${\bf A} ^{(0)} _{\pm}$ is the same. Due to the addition of the $e^{-mr}$ factor in ${\bf A} _\pm$ these two 3-vector potentials are not related by a gauge transformation and thus the two resulting ${\bf B}$ -fields are distinct. \\
 
(iii) Since the string singularities in ${\bf B} _\pm$ are real, one can no longer hide them for {\it isolated} magnetic charges. One can still effectively hide these infinite string singularities in ${\bf B}_\pm$ by requiring that magnetic charges always come in permanently confined $+g$ and $-g$ pairs ({\it i.e.} monopole-antimonopole pairs). These opposite magnetic charge pairs would still have a {\it finite}, singular string connecting them. This is like a classical version of the confinement of color charge in QCD, where quarks are permanently confined via a finite string ({\it i.e.} a color field flux tube) running between the quarks. Another option for smoothing out the string singularities in the magnetic field of the present solutions might be to introduce the photon mass via a symmetry breaking scalar field, rather than putting it in by hand as is done here. This smoothing of field singularities via the introduction of a symmetry breaking scalar field is exactly what occurs for 't Hooft-Polyakov \cite{polyakov} monopoles, where the point singularity of a Dirac monopole is smoothed into a finite core.  \\
 
(iv) The magnetic field in the present case is no longer spherically symmetric as can be seen explicitly by comparing the magnetic field from the massive photon case as given in \eqref{b-yukawa} with the Coulomb magnetic field ${\bf B} = g \frac{{\bf {\hat r}}}{r^2}$. In the presence of magnetic charge there is a connection between the gauge charge, $e$, and the space-time symmetry of rotational invariance connected with angular momentum as given in \eqref{ang3d}. Given this connection between the gauge charge, $e$, and angular momentum/rotational invariance one might suspect that rotational invariance will be broken if the gauge symmetry is broken by adding a photon mass.

\section{Angular momentum and Dirac quantization condition}

In this section we investigate the Dirac quantization condition in the presence of a photon mass. We follow the field angular momentum derivation of the Dirac quantization condition: calculate the electromagnetic field angular momentum of a electric charge-magnetic charge system and then require this to be some integer multiple of $\frac{\hbar}{2}$ \cite{saha} \cite{saha1} \cite{wilson}. A rigorous justification for this procedure in terms of the commutation relationships of the total angular momentum (mechanical plus field angular momentum) is given in \cite{fierz,lipkin,yang}. In the present work we will consider the simplified case when the electric charge and magnetic charge are at the same location. For a massless photon, if one put the electric charge and magnetic charge at the same location, one arrives at the trivial result of zero field angular momentum. However, with a massive photon the magnetic field of the monopole has a ${\bf {\hat \theta}}$ component (see \eqref{b-yukawa-1}) which gives a non-zero field angular momentum even when charges are at the same location. The more difficult case when the two charges are at different locations is work in progress \cite{evans}. 

There are several changes one must take into account when one has a massive photon. First, the energy-momentum tensor has an additional term compared to the massless photon case. For a massless photon one has $T^{0i} = ({\bf E} \times {\bf B})^i$ while for the massive case one has $T^{0i} = ({\bf E} \times {\bf B})^i + m^2 \phi_e {\bf A} ^i$ (see chapter 12 of \cite{jackson}, in particular problem 12.16). This additional term, $m^2 \phi_e {\bf A} ^i$, gives rise to an extra term in the angular momentum density. Using the fields from \eqref{e-yukawa-1} \eqref{b-yukawa-1} and the potentials from \eqref{e-yukawa} \eqref{b-yukawa} we find the field angular momentum density is
\begin{equation}
\label{ang-mom-den}
{\bf r} \times ({\bf E} \times {\bf B}_\pm) + m^2 \phi_e {\bf r} \times {\bf A}_\pm  = - e g m (1 + mr) \frac{e^{-2 m r}}{r^2} \left[\frac{\pm 1 - \cos \theta }{\sin \theta} \right] {\bf {\hat \theta}} -  eg m^2 \frac{e^{-2mr}}{r} \left[\frac{\pm 1 - \cos \theta }{\sin \theta} \right] {\bf {\hat \theta}}~,
\end{equation}
where we have used ${\bf {\hat r}} \times {\bf {\hat r}} =0$, ${\bf {\hat r}} \times {\bf {\hat \theta}} = {\bf {\hat \varphi}}$ and
${\bf {\hat r}} \times {\bf {\hat \varphi}} = - {\bf {\hat \theta}}$. Recalling that ${\bf {\hat \theta}}= \cos \theta \cos \varphi {\bf {\hat x}}+ \cos \theta \sin \varphi {\bf {\hat y}} -  \sin \theta {\bf {\hat z}}$ the $d \varphi$ integration of the volume integral of ${\bf r} \times ({\bf E} \times {\bf B}) + m^2 \phi_e {\bf r} \times {\bf A}_\pm$ will give zero for the ${\bf {\hat x}}, {\bf {\hat y}}$ components and $2 \pi$ for the ${\bf {\hat z}}$. Preforming the remaining integrals in $dr$ and $d \theta$ (which we change to $dx$ via the standard substitution $x=\cos \theta$) gives
\begin{eqnarray}
\label{ang-mom}
\frac{1}{4 \pi} \left( \int {\bf r} \times ({\bf E} \times {\bf B}_\pm) + m^2 \phi_e {\bf r} \times {\bf A}_\pm  \right) d^3 x 
= \frac{e g m}{2} \int _0 ^\infty (1 + 2 m r) e^{-2 m r} dr \int _{-1} ^{1} ( \pm 1 - x ) d x {\bf {\hat z}} 
= \pm eg {\bf {\hat z}}  ~.
\end{eqnarray}
The result (in this special case) is exactly the same as in massless photon case! Applying the condition that this field angular momentum in \eqref{ang-mom} equal $n \frac{\hbar}{2}$ thus yields the same Dirac condition on the charges as for a massless photon. This is supported by recent work \cite{goldhaber} which argues that adding both magnetic charge and photon mass to electromagnetism (mathematically the system given by equations \eqref{e-yukawa} \eqref{e-yukawa-1} \eqref{b-yukawa} \eqref{b-yukawa-1}) one should still obtain the usual Dirac quantization condition.  

One might wonder why the photon, mass $m$ does not appear in the field angular momentum result in \eqref{ang-mom}. The non-appearance of $m$ is connected with the special case we are considering, where both $e$ and $g$ are at the origin {\it i.e.} we are working in the limit $r \to 0$. In this limit the effect of the exponential factor in the fields and potentials is not felt since $e^{-mr} \to 1$. In the case when the electric charge and magnetic charge are not placed at the same location we have found \cite{evans} that the photon mass, $m$, does enter into the expression for the field angular momentum. In the case when electric charge is displaced along the $z$-axis to a point $\pm R {\hat {\bf z}}$, and the string singularity of the magnetic field runs through the electric charge (for the electric charge at $ \pm R {\hat {\bf z}}$ this means the ${\bf B}_\mp$ magnetic field) we find that the field angular momentum is ${\bf L}_{EM} = \mp 2 eg \exp[-m R]$. In the case when the string singularity of the magnetic field does not pass through the electric charge (for the electric charge at $ \pm R {\hat {\bf z}}$ this means the ${\bf B}_\pm$ magnetic field) we find that the field angular momentum is ${\bf L}_{EM} = 0$. These two preliminary results with the electric charge moved to some place on the $z$-axis, are in agreement with the result for the field angular momentum with both charges at the origin. In equation \eqref{ang-mom} the field angular momentum comes entirely from the string singularity part of the magnetic field (the ${\hat \theta}$ term in \eqref{b-yukawa-1}), and the point singularity part of the magnetic field (the ${\hat {\bf r}}$ term in \eqref{b-yukawa-1}) contributes nothing. The non-contribution of the ${\hat {\bf r}}$ term when the two charges are at the origin is exactly what one finds in the massless photon case. When the electric charge is moved to $\pm R {\hat {\bf z}}$, and with the magnetic field string singularity running through the electric charge, one has equal contributions from both the ${\hat {\bf r}}$ and ${\hat \theta}$ terms of the magnetic field from \eqref{b-yukawa-1}, but now both contributions are reduced by the expected $\exp[-m R]$ factor. When the magnetic field string singularity does not run through the electric charge, one has equal magnitude but opposite sign contributions from the ${\hat {\bf r}}$ and ${\hat \theta}$ terms of the magnetic field, and so one gets $L_{EM} =0$. The general case for the electric and magnetic charges at different locations is much more involved and we are currently working out the field angular momentum for this general case. We will then use this result to study the quantum aspects of magnetic charge plus photon mass by working out how the total angular momentum commutation relationships work out, as is done for the massless photon case in references \cite{fierz,lipkin,yang}. We hope to report on results of this much more complex analysis soon \cite{evans}. \\

\section{Summary and Discussion}

In this work we have analyzed how the theory of magnetic charge is altered by the inclusion of a photon mass. The system of equations used ({\it i.e.}  \eqref{e-maxwell3-source} \eqref{e-maxwell3-source-1} \eqref{b-maxwell3-source} \eqref{b-maxwell3-source-1}) were those considered in reference \cite{joshi}. In this paper, as in \cite{joshi}, the photon mass term is put in by hand. More realistically we have in mind that the photon mass term should arise from the Higgs mechanism which would mean introducing a symmetry breaking scalar field. The introduction of a dynamical Higgs-like scalar field might connect with the approach used reference \cite{guimaraes} which studied magnetic charge plus photon mass using a generalized Julia-Toulouse approach to handle topological currents. This resulted in a low energy effective theory where magnetic charges where confined by open, non-singular magnetic vortices.    

The differences we found from the case of magnetic charge  with a massless photon are: (i) The string is now a real singularity which shows up, not only in the 3-vector potential, but in the magnetic field  -- see equations \eqref{A-string} \eqref{b-yukawa} \eqref{b-yukawa-1}. (ii) The magnetic field has a ${\bf {\hat \theta}}$-component in addition to the expected ${\bf {\hat r}}$-component -- see equation \eqref{b-yukawa-1}. The spherical symmetry of the magnetic field is lost. (iii) The two different 3-vector potentials from equation \eqref{b-yukawa} are distinct, gauge inequivalent solutions, which give different magnetic fields. In the massless photon case the 3-vector potentials ${\bf A}^{(0)}_\pm$, are gauge equivalent and give the same magnetic field. (iv) As in the massless photon case the system of a magnetic charge plus electric charge carries a field angular momentum. Unlike the massless photon case, this field angular momentum persisted when the two charges are at the same location due to the new ${\bf {\hat \theta}}$-component of the magnetic field in the massive photon case. This ${\bf {\hat \theta}}$-component is connected with the string singularity in the magnetic field. (v) In the special case when the electric and magnetic charges are at the same location, the field angular momentum has a magnitude $eg$, which is exactly the same as in the massless case. Applying the requirement that this field angular momentum equal some integer multiple of $\frac{\hbar}{2}$ then returns the usual Dirac condition \eqref{dirac-cond}. 

One of the worrisome aspects of the present solutions is the real string singularity coming from the ${\hat \theta}$ component of the magnetic field in \eqref{b-yukawa-1}. One way to ``cut" this infinite, singular string down to a finite, singular string is to propose that magnetic charges are always confined into $+g$ and $-g$ pairs, with string singularities running in the same direction. For example, by placing a $+g$ charge at ${\bf r}=0$ and a $-g$ charge at ${\bf r} =  R {\bf {\hat z}}$, with each charge having the field ${\bf B}_\pm ({\bf r})$ and ${\bf B}_\pm ({\bf r} - R{\bf {\hat z}})$ respectively, one will get a partial cancellation of the strings from each charge in the region $-\infty < z < 0 $ for the $B_+$ case and in the region $R < z < \infty$ for the $B_-$ case. In the region $0<z<R$ one still has a finite, singular string connecting the magnetic charges. This is similar to the flux tube model of QCD confinement where color electric charges are confined by a color electric flux tube stretching between the charges. Reference \cite{ahrens} comes to a similar conclusion: that magnetic charges are confined into oppositely charged pairs connected by a finite length (but singular) Dirac string when the photon is massive. The confinement of magnetic charge with a massive photon was also found in reference \cite{guimaraes} which used a generalized Julia-Toulouse approach \cite{julia} \cite{quevedo} \cite{grigorio} for handling the condensation of topological currents (charges or defects). One difference between the string confinement picture of the present paper (and also of reference \cite{ahrens}), and the QCD flux tube confinement picture or the results from reference \cite{guimaraes}, is that the finite Dirac strings connecting the magnetic charges in this work (or in \cite{ahrens}) are singular, while the flux tubes in QCD or the magnetic vortex of \cite{guimaraes} are non-singular. This is to be expected since the solutions presented here and those given in \cite{ahrens} are purely classical. In QCD the flux tubes arise in the context of a quantized theory and thus one might hope that quantizing the present solutions might turn the singular classical strings into non-singular, quantum strings. 

Since photons have a very low upper bound to their mass ($m_{photon} < 10^{-18}$ eV \cite{pdg}) one might suppose that their mass is exactly zero and so the above analysis has no physical application. However, there are some physical quantities which are ``anomalously small" yet non-zero ({\it e.g.} the cosmological constant) so one should keep open the option the that photon mass is small but non-zero. In addition one could think to embed the Abelian, $U(1)$ monopole string solutions of the present work in a non-Abelian theory where the ``photons" do have a mass or an effective mass. In the electroweak interaction one has a massive ``photon" (the $Z$-boson); in QCD gluons can be thought of as having an effective mass.        

\section*{Acknowledgment}

DS is supported by grant $\Phi.0755$ in fundamental research in Natural Sciences by the Ministry of Education and Science of Kazakhstan. We acknowledge the helpful comments of an anonymous referee and especially the discussions with Gerardo Mu{\~n}oz, which helped to fix some errors in the original manuscript.

\end{document}